\documentclass[journal]{IEEEtran}

\usepackage{cite}

\usepackage{amsmath}
\usepackage{mathrsfs}

\usepackage{algorithmic}

\usepackage{array}

\usepackage{graphicx}
\usepackage{subfigure} 
\usepackage{caption}
\usepackage{url}
\usepackage{graphicx,amsmath,amssymb,amsfonts}
\usepackage{algorithmic,algorithm}

\usepackage{xcolor}

\newtheorem{theorem}{\textbf{Theorem}}

\newtheorem{lemma}{\textbf{Lemma}}

\newtheorem{corollary}{\textbf{Corollary}}

\makeatletter

\newcommand{\Rmnum}[1]{\expandafter\@slowromancap\romannumeral #1@}
\makeatother

\begin{document}

\title{
	{\fontsize{19.0 pt}{\baselineskip}\selectfont Intelligent Reflecting Surface Aided Secure UAV Communications} % <-this is double column
}

\author{Wen~Wang, ~Hui~Tian, ~Wanli~Ni, and ~Meihui~Hua
}
	
\maketitle

\begin{abstract}
	In this letter, we study the secure communication problem in the unmanned aerial vehicle (UAV) enabled networks aided by an intelligent reflecting surface (IRS) from the physical-layer security perspective.
	Specifically, the IRS is deployed to assist the wireless transmission from the UAV to the ground user in the presence of an eavesdropper.
	The objective of this work is to maximize the secrecy rate by jointly optimizing the phase shifts at the IRS as well as the transmit power and location of the UAV.
	However, the formulated problem is difficult to solve directly due to the non-linear and non-convex objective function and constraints.
	By invoking fractional programming and successive convex approximation techniques, the original problem is decomposed into three subproblems, which are then transformed into convex ones. Next, a low-complexity alternating algorithm is proposed to solve the challenging non-convex problem effectively, where the closed-form expressions for transmit power and phase shifts are obtained at each iteration.
	Simulations results demonstrate that the designed algorithm for IRS-aided UAV communications can achieve higher secrecy rate than benchmarks.
\end{abstract}


\section{Introduction}
By manipulating the transmit signal to maximize the secure rate in the wireless networks, the physical layer security (PLS) technique is able to effectively reduce the information leakage and enhance data security \cite{PLS-1, PLS-2}.
However, the conventional scheme of implementing PLS in the ground base station (BS) is severely limited to the location of legitimate users and eavesdroppers \cite{UAV-1}.
Due to the high mobility and flexible deployment of unmanned aerial vehicles (UAVs), they are capable of offering strong line-of-sight (LoS) links and adjusting transmit strategy dynamically to boost the security \cite{UAV-2, UAV-3}, which makes the combination of UAV and PLS an appealing scheme to provide ubiquitous secure wireless service in the fifth generation and beyond.
Whereas, this approach surmounts the location restrictions on conventional static ground scenes at the cost of the LoS links also being leveraged by eavesdroppers, which greatly increases the potential risk of security in UAV communications \cite{UAV-4, UAV-5, UAV-6, UAV-7}. 

Against the above background and issues, an intuitive idea is to establish a more controllable wireless environment to improve the secure performance of UAV communication systems.
Recently, intelligent reflecting surface (IRS) has emerged as a revolutionary technique due its ability of reshaping wireless channels \cite{IRS-1}, which adds more degrees of freedom to achieve a smart and reconfigurable wireless environment in a controllable manner.
Technically, these tunable and low-cost reflecting elements equipped on IRS are capable of dynamically adjusting the phase shifts and absorbing the signal energy, then the desired signals can be boosted yet the interference signals are diminished simultaneously \cite{IRS-2}.
Therefore, IRS can be carefully designed to improve the undesirable propagation conditions to facilitate UAV communications.
However, integrating IRS into UAV-enabled secure communication systems faces challenges from network characterization to performance optimization \cite{challenge}.

\begin{figure}[!t]
	\centering
	\includegraphics[width=3.2 in]{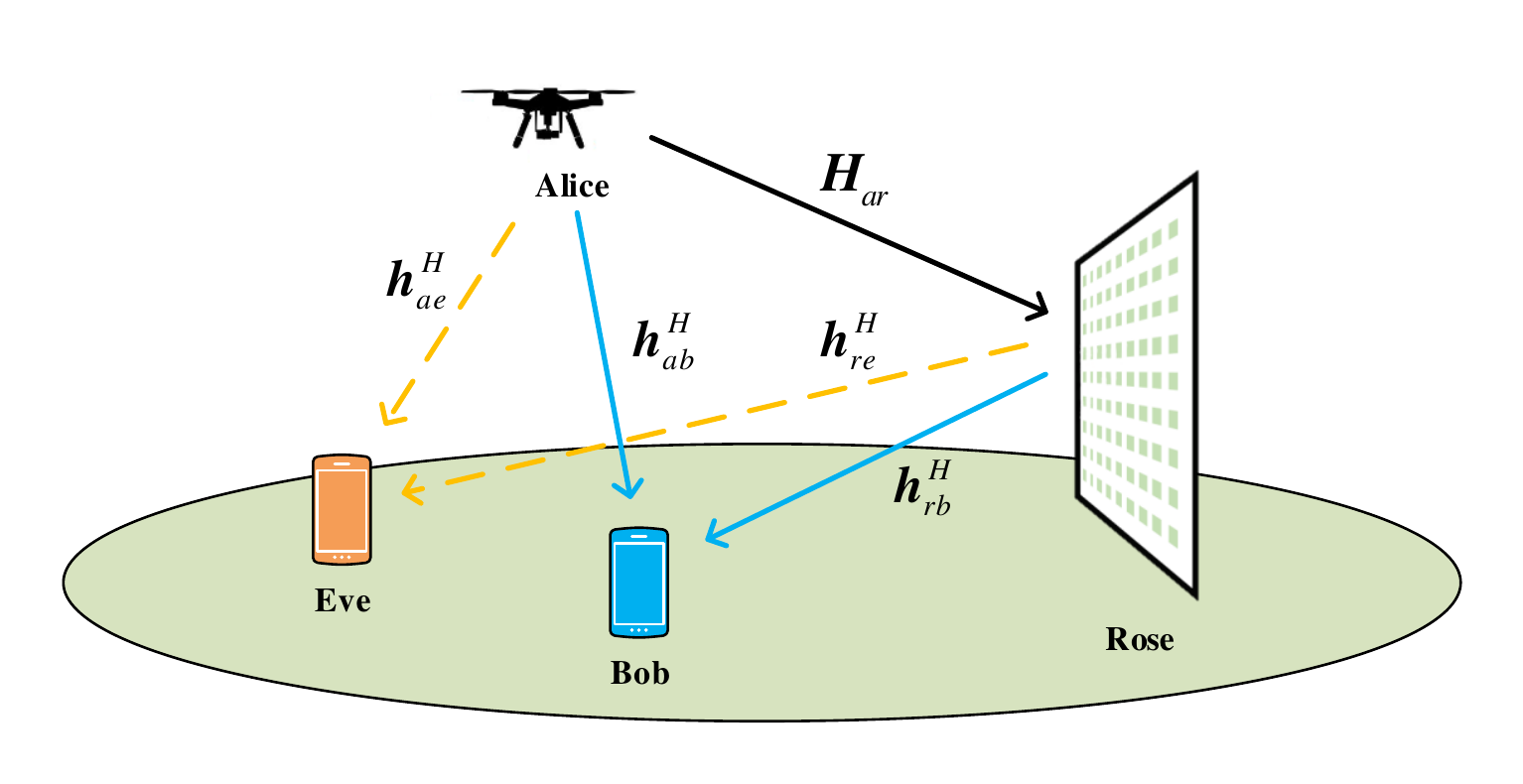}
	\caption{System model of IRS-aided UAV communications.}
	\label{model}
\end{figure}

Inspired by these great challenges and potentials of UAV and IRS in enhancing PLS, the secrecy performance of both UAV-enabled and IRS-aided communications has received considerable attention. 
For example, assuming that the position of the eavesdropper was available under LoS propagations, the trajectory and transmit power of the UAV were jointly optimized to maximize the secrecy rate in \cite{UAV-1, UAV-2}.
In addition, authors in \cite{UAV-3, UAV-4} proposed the transmit jamming strategy to secure wireless communication in which UAVs send jamming signals to confound the eavesdroppers.
However, the broadcast nature of wireless transmissions and the dominating LoS channels bring more eavesdroppping threats in such UAV deployed scenarios, and the incorporation of the cooperative jamming inevitably incurs synchronization and 
communication overhead for transmit power control.
On the other hand, by leveraging the fine-grained tunable capability of IRSs, the works in \cite{IRS-3, 8743496, CSI_1, CSI_2} studied the secrecy performance of IRS assisted communication systems.
By jointly designing the active and passive beamforming, the achievable secrecy rate can be significantly improved. 
In particular, authors in \cite{IRS} provided an overview of the secure wireless communications aided by IRS, including its classification, challenges and recommendations. 
However, most of the existing contributions on IRS-aided PLS are restricted to the static ground scenes with fixed terrestrial facilities. 
To the best of our knowledge, the study on UAV-enabled secure communication is still an open issue and the corresponding numerical results or analysis are insufficient, not to mention the integration of IRS.

Different from the conventional terrestrial secure communication, this letter proposes a novel secrecy rate maximization problem by exploiting both the high mobility UAV and the software controllable IRS for further secure performance enhancement. 
More specifically, the contributions of this work can be summarized as follows:
1) we design a novel framework for UAV-enabled secure communications aided by IRS, where a non-linear and non-convex problem is formulated to maximize the secrecy rate;
2) we propose a low complexity algorithm to solve this challenging problem, where the closed-form solutions for transmit power and phase shifts are obtained at each iteration;
3) we conduct numerical simulations to demonstrate the impacts of maximum transmit power, the number of reflecting elements and the distance between IRS and the legitimate user, and all of them validate the performance gain of the proposed algorithm.

\section{System Model and Problem Formulation}

\subsection{System Model}
As shown in Fig.\ref{model}, we consider an IRS-aided UAV network where one IRS (Rose) is deployed to assist the secure communication from the UAV (Alice) to a legitimate ground user (Bob), against an eavesdropper (Eve).
Specifically, Rose consists of $M$ passive reflecting elements, Alice is equipped with $N$ antennas, and both Bob and Eve are single-antenna users.
Assume that the height of Bob and Eve is zero, while that of Alice and Rose are $H$ and $h$, respectively.
Meanwhile, Alice, Bob, Eve, and Rose are horizontally located at $\boldsymbol{a}$, $\boldsymbol{b}$, $\boldsymbol{e}$, $\boldsymbol{r}$, and they are indexed by $a$, $b$, $e$, $r$, respectively. 

Let $\boldsymbol{h}_{ab}\in {\mathbb{C}^{N}}$, $\boldsymbol{h}_{ae}\in {\mathbb{C}^{N}}$, $\boldsymbol{H}_{ar}\in {\mathbb{C}^{M\times N}}$, $\boldsymbol{h}_{rb}\in {\mathbb{C}^{M}}$, and $\boldsymbol{h}_{re}\in {\mathbb{C}^{M}}$ denote the channel gain\footnote{It is worth mentioning that, by applying the existing channel estimation method in \cite{CSI_2}, we assume that the global channel state information knowledge is available at both Alice and Rose.} of the Alice-Bob, Alice-Eve, Alice-Rose, Rose-Bob, and Rose-Eve links, respectively.
Then, the channel model\footnote{Since both UAV and IRS are deployed at high altitude, and the complex environment has rich scattering, all channels in the system are likely to establish LoS links and also experience small-scale fading \cite{UAV-4}.} can be expressed as
$\boldsymbol{h}_{mn}^H=\sqrt{\beta_0d_{mn}^{-c_{mn}}}\boldsymbol{g}_{mn}, m\in\{a,r\}, n\in\{b,e\}$,
where $\beta_0$ is the power gain at the reference distance of 1 m, $d_{mn}$ denotes the distance between $m$ and $n$, $c_{mn}$ is the path loss factor.
Moreover, the small-scale fading $\boldsymbol{g}_{mn}$ can be modeled as
\begin{eqnarray}
\label{1}
\boldsymbol{g}_{mn}=\sqrt{\frac{k_{mn}}{k_{mn}+1}}\boldsymbol{g}+\sqrt{\frac{1}{k_{mn}+1}}\widetilde{\boldsymbol{g}}, 
\end{eqnarray}
where $\boldsymbol{g}$ represents the deterministic LoS channel component, $\widetilde{\boldsymbol{g}}$ denotes the random scattered component, and $k_{mn}$ is the Rician factor\footnote{The Rician factor is given by $k_{mn}=A_1\exp(A_2\theta_{mn})$, where $A_1$ and $A_2$ are constant coefficients determined by the specific environment, and $\theta_{mn}=\arcsin(\frac{\bar{h}}{d_{mn}})$ is the elevation angle between the higher transmitter $m$ and the relatively low user $n$, with $\bar{h}$ denoting their height difference.}.
As such, the channel gain of Alice-Rose link can be obtained as $\boldsymbol{H}_{ar}=\sqrt{\beta_0d_{ar}^{-c_{ar}}}(\sqrt{\frac{k_{ar}}{k_{ar}+1}}\boldsymbol{G}+\sqrt{\frac{1}{k_{ar}+1}}\widetilde{\boldsymbol{G}})$, where $\boldsymbol{G}$ and $\widetilde{\boldsymbol{G}}$ are similar to $\boldsymbol{g}$ and $\widetilde{\boldsymbol{g}}$, respectively.

Thus, the received signal at Bob and Eve can be given by
\begin{eqnarray}
\label{2}
s_n= (\boldsymbol{h}_{rn}^H\boldsymbol{\Theta}\boldsymbol{H}_{ar} +\boldsymbol{h}_{an}^H)\boldsymbol{f}x+\xi_n, n\in \{b,e\},
\end{eqnarray}
where $\boldsymbol{f}\in\mathbb{C}^{N}$ is the transmit precoding vector, $x$ is the information-bearing symbol of the legitimate receiver Bob with $E [|x|^2 ]=1$, $\boldsymbol{\Theta}=\textrm{diag}(e^{j\phi_1},\cdots,e^{j\phi_m},\cdots,e^{j\phi_M})$ is the diagonal matrix with $\phi_i\in[0,2\pi)$ denoting the phase shift incurred by the $i$-th reflecting element of Rose,  $\xi_b\sim\mathcal{N}(0,\sigma_{b}^2)$ and $\xi_e\sim\mathcal{N}(0,\sigma_{e}^2)$ represent the additive white Gaussian noise at Bob and Eve, respectively.

Then, the signal-to-interference-plus-noise ratio at Bob and Eve can be expressed as
\begin{eqnarray}
\label{3}
\gamma_n=\frac{\left|\left(\boldsymbol{h}_{rn}^H\boldsymbol{\Theta}\boldsymbol{H}_{ar}+\boldsymbol{h}_{an}^H\right)\boldsymbol{f}\right|^2}{\sigma_n^2}, n \in\{b,e\}.
\end{eqnarray}

\subsection{Problem Formulation}
By jointly designing the transmit power and location at Alice as well as the phase shifts at Rose, the objective of this work is to maximize the system secrecy rate.
Mathematically, the achievable secrecy rate can be given by
\begin{eqnarray}
\label{4}
R_s\left(\boldsymbol{f},\boldsymbol{\Theta},\boldsymbol{a}\right) = \left[ \log_2 \left( 1 + \gamma_{b} \right)  - \log_2 \left( 1 + \gamma_{e} \right) \right]^+,
\end{eqnarray}
where $[x]^+ = \max \{x,0\}$ and we omit this operator $[\cdot]^+$ in the rest of this work for simplicity, due to the non-negative nature of the optimal secrecy rate.
Then, the secrecy rate maximization problem can be formulated as
\begin{subequations} \label{5}
	\begin{eqnarray}
	\label{5a}
	&\underset{\boldsymbol{f},\boldsymbol{\Theta},\boldsymbol{a}}{\rm max}&\ {R_s\left(\boldsymbol{f},\boldsymbol{\Theta},\boldsymbol{a}\right)} \\
	\label{5b}
	&\operatorname{s.t.}&{\lVert\boldsymbol{f}\lVert}^2\leq P_\textrm{max}, \\
	\label{5c}
	&&\phi_m\in[0,2\pi), \ m=1,\cdots M, \\
	\label{5d}
	&&\log_2 \left(1+\gamma_{b}\right)\geq R_\textrm{min},
	\end{eqnarray}
\end{subequations}
where the constraint (\ref{5b}) guarantees that the maximum transmit power at Alice is no more than the budget $P_\textrm{max}$, (\ref{5c}) is the phase shifts constraint at the Rose, and (\ref{5d}) ensures that the minimum rate requirement $R_\textrm{min}$ is satisfied at Bob.

Note that problem (\ref{5}) is non-trivial to solve due to the non-convex objective function and unit-modulus constraints, as well as the coupled variables $\boldsymbol{f}$, $\boldsymbol{\Theta}$ and $\boldsymbol{a}$.
To solve this challenging problem, we first propose to decompose it into tractable subproblems, then an alternating optimization (AO) algorithm is designed to address the original problem (\ref{5}) effectively, i.e., fix one and optimize the other, then repeat this in turn until the convergence criterion is reached.

\section{Alternating Optimization}
\subsection{Power Allocation}
In this section, we first investigate the optimization of $\boldsymbol{f}$ with the fixed $\boldsymbol{\Theta}$ and $\boldsymbol{a}$. 
By introducing the matrix variable $\boldsymbol{Q}_n=(\boldsymbol{h}_{rn}^H\boldsymbol{\Theta}\boldsymbol{H}_{ar}+\boldsymbol{h}_{an}^H)^H(\boldsymbol{h}_{rn}^H\boldsymbol{\Theta}\boldsymbol{H}_{ar}+\boldsymbol{h}_{an}^H), n \in\{b,e\}$, problem (\ref{5}) can be reformulated as
\begin{subequations}\label{7}
	\begin{eqnarray}
	&\underset{\boldsymbol{f}}{\rm max}&\
	{\frac{{\boldsymbol{f}}^H\boldsymbol{Q}_b\boldsymbol{f}+\sigma_{b}^2}{{\boldsymbol{f}}^H\boldsymbol{Q}_e\boldsymbol{f}+\sigma_{e}^2}} \\
	&\operatorname{s.t.}&{\lVert \boldsymbol{f}\lVert}^2 \leq P_\textrm{max}, \\
	&&{\boldsymbol{f}}^H\boldsymbol{Q}_b\boldsymbol{f}\geq\left(2^{R_\textrm{min}}-1\right)\sigma_{b}^2.
	\end{eqnarray}
\end{subequations}

To ensure that problem (\ref{7}) has feasible solutions, we assume $P_\textrm{max}\geq \frac{(2^{R_\textrm{min}}-1)\sigma_{b}^2}{\lVert \boldsymbol{Q}_b\lVert}$. 
Therefore, the optimal solution to the generalized eigenvalue problem (\ref{7}) can be derived as \cite{IRS-2}
\begin{eqnarray}
\label{8}
&\boldsymbol{f}_\textrm{opt}=\sqrt{P_\textrm{max}}\boldsymbol{e}_\textrm{max},
\end{eqnarray}
where $\boldsymbol{e}_\textrm{max}$ is the normalized dominant generalized eigenvector of the matrix pencil $(P_\textrm{max}\boldsymbol{Q}_b+\sigma_{b}^2\boldsymbol{I}, P_\textrm{max}\boldsymbol{Q}_e+\sigma_{e}^2\boldsymbol{I})$,
where $\boldsymbol{I}$ denotes the identity matrix.

\subsection{Reflection Design}
Next, with given $\boldsymbol{f}$ and $\boldsymbol{a}$, we define $\boldsymbol{h}_B=\textrm{diag}(\boldsymbol{h}_{rb}^H)\boldsymbol{H}_{ar}\boldsymbol{f}$, $\widetilde{h}_B=\boldsymbol{h}_{ab}^H\boldsymbol{f}$, $\boldsymbol{h}_E=\textrm{diag}(\boldsymbol{h}_{re}^H)\boldsymbol{H}_{ar}\boldsymbol{f}$, and $\widetilde{h}_E=\boldsymbol{h}_{ae}^H\boldsymbol{f}$. Then, by invoking the equality $\boldsymbol{h}\boldsymbol{\Theta}\boldsymbol{b}=\boldsymbol{\theta}^H\textrm{diag}(\boldsymbol{h})\boldsymbol{b}$ where $\boldsymbol{\theta}=[e^{j\phi_1},\cdots,e^{j\phi_M}]^H$, 
problem (\ref{5}) can be rewritten as
\begin{subequations}\label{9}
	\begin{eqnarray}
	\label{9a}
    &\underset{\boldsymbol{\theta}}{\rm max}&
    {\frac{\left|\boldsymbol{\theta}^H\boldsymbol{h}_B+\widetilde{h}_B\right|^2+\sigma_{b}^2}{\left|\boldsymbol{\theta}^H\boldsymbol{h}_E+\widetilde{h}_E\right|^2+\sigma_{e}^2}}  \\
    &\operatorname{s.t.}&\phi_m\in[0,2\pi), \ m=1,\cdots M, \\
    &&R\left(\boldsymbol{\theta}\right)\geq0,
	\end{eqnarray}
\end{subequations}
where $R(\boldsymbol{\theta})=|\boldsymbol{\theta}^H\boldsymbol{h}_B+\widetilde{h}_B|^2+\sigma_{b}^2{-2}^{R_\textrm{min}}\sigma_{b}^2$.

It can be noticed that problem (\ref{9}) is a fractional programming problem \cite{Werner1967On}. As such,
by adding a non-negative multiplication factor $\mu$, we transform problem (\ref{9}) into the following approximate problem
\begin{subequations} \label{10}
	\begin{eqnarray}
	\label{10a}
	&\underset{\boldsymbol{\theta}}{\rm min}&
	\varphi\left(\boldsymbol{\theta}|\mu\right) \\
	&\operatorname{s.t.}&\left(\textrm{8b}\right) \ {\rm and} \ \left(\textrm{8c}\right),
	\end{eqnarray}
\end{subequations}
where $\varphi(\boldsymbol{\theta}|\mu)=|\boldsymbol{\theta}^H\boldsymbol{h}_E+\widetilde{h}_E|^2+\sigma_{e}^2-\mu R(\boldsymbol{\theta})$.
Denote $\varphi^\star(\boldsymbol{\theta}|\mu)$ as the optimal objective value of problem (\ref{10}), then the optimal objective value of problem (\ref{9}) is equivalent to the unique root of $\varphi^\star(\boldsymbol{\theta}|\mu)=0$ \cite{Werner1967On}. 
However, obtaining the root requires solving problem (\ref{10}) with given $\mu$, which is still a complicated non-convex problem with unit-modulus constraints.

To make it more tractable, we follow the second-order Taylor expansion method \cite{7547360} to minimize an upper bound of the objective function (\ref{9a}), which can be given by
\begin{eqnarray}
\label{11}
\varphi\left(\boldsymbol{\theta}|\mu\right)
&\le&\lambda_\textrm{max}\left(\boldsymbol{H}\right)\lVert\boldsymbol{\theta}\lVert^2-2\boldsymbol{\Re}\left\{\boldsymbol{\theta}^H{\boldsymbol{\beta}}\right\}+c,
\end{eqnarray}
where
\begin{subequations}
\begin{eqnarray}
	\label{12}
	\boldsymbol{H} &=&\boldsymbol{h}_E\boldsymbol{h}_E^H-\mu\boldsymbol{h}_B\boldsymbol{h}_B^H, \\
	\boldsymbol{\beta} &=& (\lambda_\textrm{max}(\boldsymbol{H})\boldsymbol{I}-\boldsymbol{H})\widetilde{\boldsymbol{\theta}}+\mu\widetilde{h}_B^\ast\boldsymbol{h}_B-\widetilde{h}_E^\ast\boldsymbol{h}_E, \\
	c &=& {\widetilde{\boldsymbol{\theta}}}^H(\lambda_\textrm{max}(\boldsymbol{H})\boldsymbol{I}-\boldsymbol{H})\widetilde{\boldsymbol{\theta}}+|{\widetilde{h}}_E|^{2}+\sigma_{e}^2 \notag \\ 
	&{}&- {\mu|{\widetilde{h}}_B|}^{2} - \mu(1-2^{R_\textrm{min}})\sigma_{b}^2.
\end{eqnarray}
\end{subequations}

Note that $\widetilde{\boldsymbol{\theta}}$ is the solution obtained at the previous iteration.
For ease of calculation, we drop the constant terms of the bound and problem (\ref{9}) can be further approximated as
\begin{subequations}\label{13}
	\begin{eqnarray}
	\label{13a}
	&\underset{\boldsymbol{\theta}}{\rm max}&
	{2\boldsymbol{\Re}\left\{\boldsymbol{\theta}^H\boldsymbol{\beta}\right\}} \\
	&\operatorname{s.t.}&\left(\textrm{8b}\right) \ {\rm and} \ \left(\textrm{8c}\right).
	\end{eqnarray}
\end{subequations}

Obviously, the objective function (\ref{13a}) is maximized only when the phases of $\boldsymbol{\theta}_i$ and $\boldsymbol{\beta}_i$ are equal. 
Thus, the closed-form solution for problem (\ref{13}) with given $\mu$ can be derived as
\begin{eqnarray}
\label{14}
\boldsymbol{\theta}^\star\left(\mu\right)=\left[e^{j\textrm{arg}\left(\beta_1\right)},\cdots,e^{j\textrm{arg}\left(\beta_M\right)}\right]^T.
\end{eqnarray}
	
By successively tuning the phase shifts of all elements based on (\ref{14}), the optimal solution to problem (\ref{10}) can be obtained as $\widetilde{\varphi}^\star(\boldsymbol{\theta}|\mu)$. 
Since $\widetilde{\varphi}^\star(\boldsymbol{\theta}|\mu)$ is a strictly decreasing function with $\widetilde{\varphi}^\star(\boldsymbol{\theta}|0)>0$ and $\widetilde{\varphi}^\star(\boldsymbol{\theta}|+\infty)<0$ \cite{8743496}, it can be verified that $\widetilde{\varphi}^\star(\boldsymbol{\theta}|\mu)=0$ has a unique root (denote by $\mu^\prime$) and $\mu^\prime\neq0$. 
Furthermore, owing to $|\boldsymbol{\theta}^H\boldsymbol{h}_E+\widetilde{h}_E|^2+\sigma_{e}^2>0$, and $\mu>0$, $R(\boldsymbol{\theta})$ is guaranteed to be non-negative when $\widetilde{\varphi}^\star(\boldsymbol{\theta}|\mu)=0$. 
As a result, the optimal phase shifts can be obtained as $\boldsymbol{\theta}^\star(\mu^\prime)$.

\subsection{UAV Deployment}
It remains to optimize $\boldsymbol{a}$ with fixed $\boldsymbol{f}$ and $\boldsymbol{\Theta}$. 
Define two constants as $\alpha_{an}=\sigma_{n}^{-1}\sqrt{\beta_0}\boldsymbol{h}_{rn}^H\boldsymbol{\Theta}\boldsymbol{f}$ and $\omega_{an}=\sigma_{n}^{-1}\sqrt{\beta_0}\boldsymbol{f}, n \in\{b,e\}$, which are independent of $\boldsymbol{a}$, then problem (\ref{5}) is reduced to
\begin{subequations}\label{15}
	\begin{eqnarray}
	\label{15a}
	&\underset{\boldsymbol{a}}{\rm max}&
	\frac{\left|\varphi\left(\boldsymbol{a},\boldsymbol{r}\right)\alpha_{ab}+\varphi\left(\boldsymbol{a},\boldsymbol{b}\right)\omega_{ab}\right|^2+1}{\left|\varphi\left(\boldsymbol{a},\boldsymbol{r}\right)\alpha_{ae}+\varphi\left(\boldsymbol{a},\boldsymbol{e}\right)\omega_{ae}\right|^2+1} \\
	&{\rm s.t.}& \left|\varphi\left(\boldsymbol{a},\boldsymbol{r}\right)\alpha_{ab}+\varphi\left(\boldsymbol{a},\boldsymbol{b}\right)\omega_{ab}\right|^2\geq2^{R_\textrm{min}}-1,
	\end{eqnarray}
\end{subequations}
where $\varphi(\boldsymbol{a},\boldsymbol{n})=d_{an}^{-\frac{c_{an}}{2}}\sqrt{(k_{an}+1)^{-1}}(\sqrt{k_{an}}\boldsymbol{g}+\widetilde{\boldsymbol{g}})$ with $d_{an}^2={\lVert\boldsymbol{a}-\boldsymbol{n}\lVert}^2+H^2, \boldsymbol{n}\in\{\boldsymbol{b},\boldsymbol{e}\}$ and similarly, $\varphi(\boldsymbol{a},\boldsymbol{r})=d_{ar}^{-\frac{c_{ar}}{2}}\sqrt{(k_{ar}+1)^{-1}}(\sqrt{k_{ar}}\boldsymbol{G}+\widetilde{\boldsymbol{G}})$ with $d_{ar}^2={\lVert\boldsymbol{a}-\boldsymbol{r}\lVert}^2+(H-h)^2$.
\begin{algorithm}[tbp]
	\caption{Alternating Optimization for Solving (\ref{5})}
	\label{Algorithm1}
	\begin{algorithmic}[1] 
		\STATE \textbf{Initialize} the maximum iteration number $L$, the current iteration number $n=0$, and the tolerance $\epsilon$;
		\REPEAT
		\STATE Given $(\boldsymbol{f}^{(n)},\boldsymbol{\Theta}^{(n)},\boldsymbol{a}^{(n)})$, calculate $\boldsymbol{f}^{(n)}$ by using (\ref{8});
		\STATE Given $(\boldsymbol{f}^{(n+1)},\boldsymbol{\Theta}^{(n)},\boldsymbol{a}^{(n)})$,
		obtain $\boldsymbol{\Theta}^{(n)}$ by applying the fractional programming method and the derived closed-form solution in (\ref{14});		
		\STATE Given $(\boldsymbol{f}^{(n+1)},\boldsymbol{\Theta}^{(n+1)},\boldsymbol{a}^{(n)})$,
		obtain $\boldsymbol{a}^{(n)}$ by applying the fractional programming method and DC algorithm;
		\STATE Update $n=n+1$;
		\UNTIL $|R_s^{(n+1)}-R_s^{(n)}|<\epsilon$ or $n>L$;
		\STATE \textbf{Output} the converged solution $(\boldsymbol{f}^{(n)},\boldsymbol{\Theta}^{(n)},\boldsymbol{a}^{(n)})$.
	\end{algorithmic}  
\end{algorithm}
Although $\varphi(\boldsymbol{a},\boldsymbol{n})$ and $\varphi(\boldsymbol{a},\boldsymbol{r})$ are neither concave nor convex w.r.t $\boldsymbol{a}$, they are respectively convex w.r.t ${\lVert\boldsymbol{a}-\boldsymbol{n}\lVert}^2+H^2$ and ${\lVert\boldsymbol{a}-\boldsymbol{r}\lVert}^2+(H-h)^2$, which allows us to leverage the successive convex approximation (SCA) technique to derive their convex approximations. 
To be specific, due to the fact that the first-order Taylor approximation of a convex function is a global under-estimator, $\varphi(\boldsymbol{a},\boldsymbol{n})$ and $\varphi(\boldsymbol{a},\boldsymbol{r})$ can be lower-bounded as follows.
\begin{figure*}[htbp]
		\begin{minipage}[t]{0.32\linewidth}
			\includegraphics[width=2.3in]{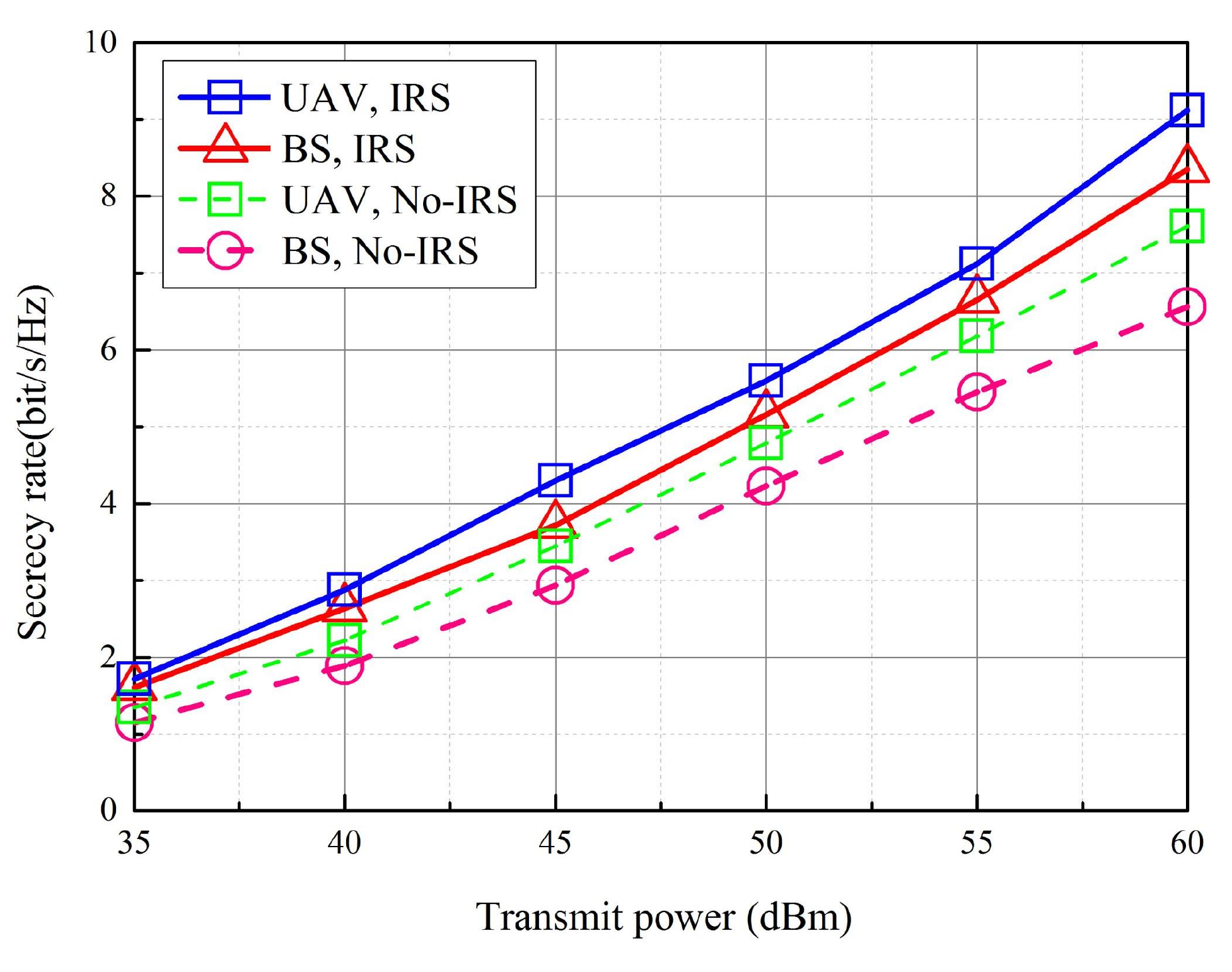}
	     	\caption{Secrecy rate vs. maximum transmit power.}
			\label{result1}
		\end{minipage}%
		\begin{minipage}[t]{0.32\linewidth}
			\includegraphics[width=2.3in]{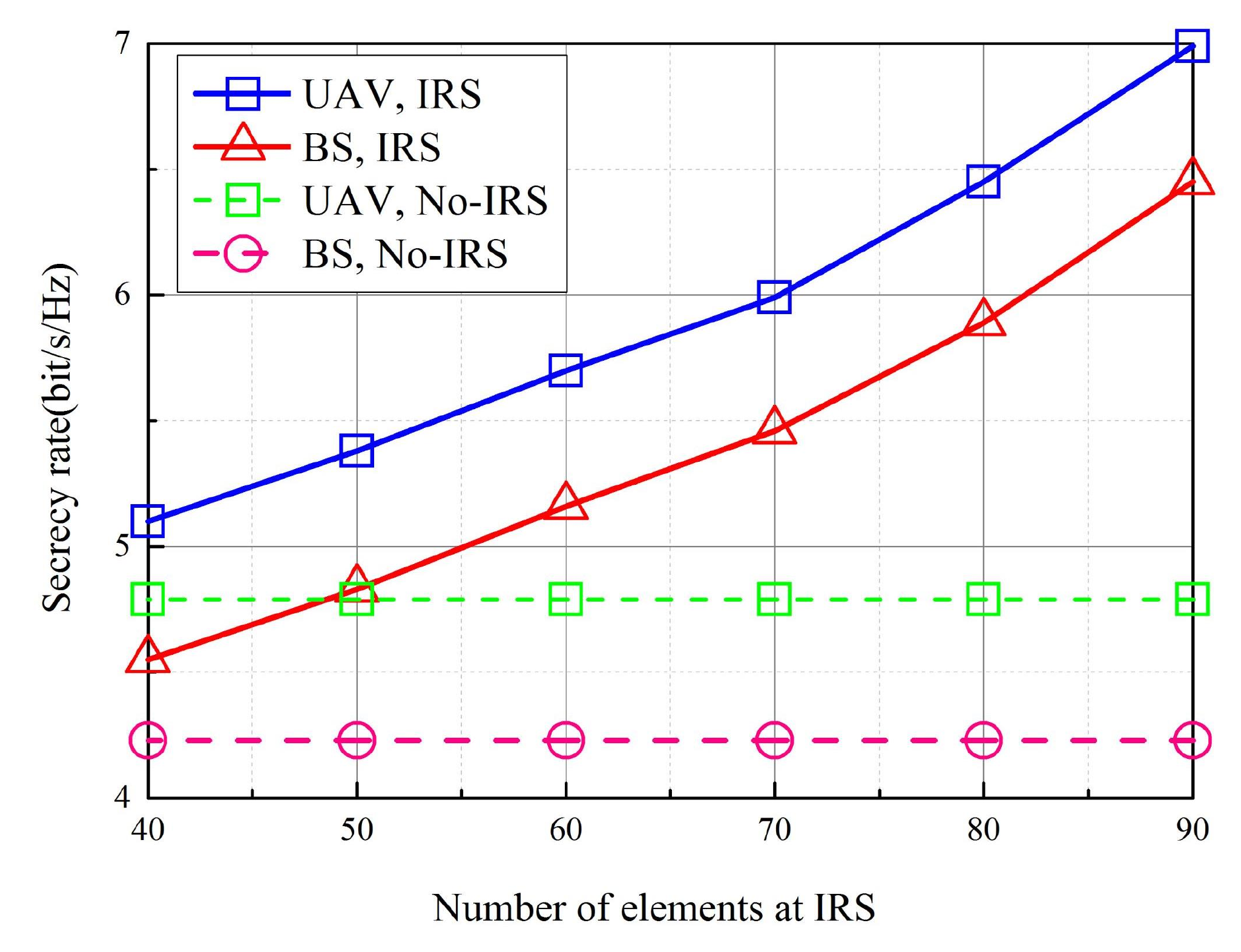}
			\caption{Secrecy rate vs. reflecting elements.}
			\label{result2}
		\end{minipage}
		\begin{minipage}[t]{0.32\linewidth}
			\includegraphics[width=2.3in]{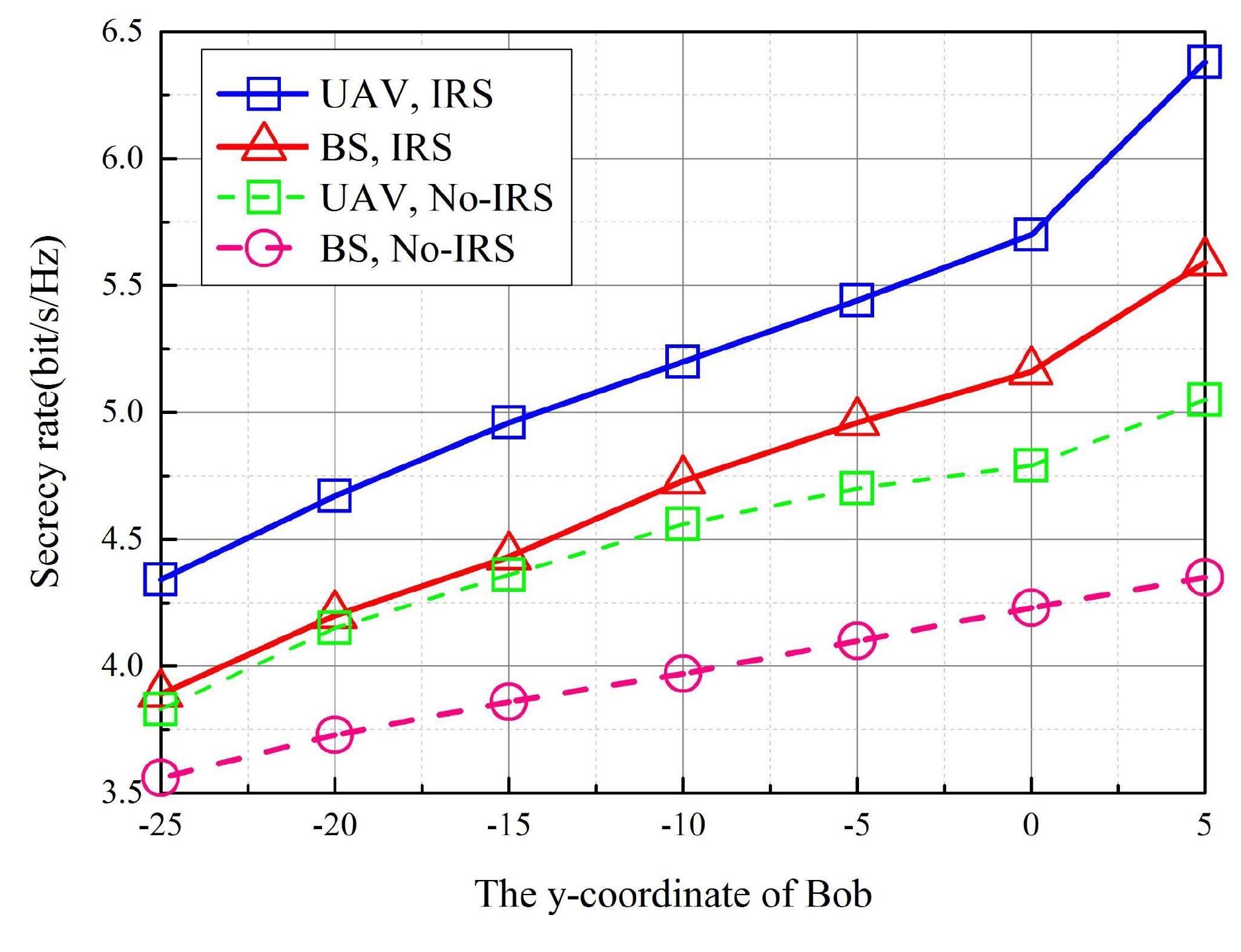}
			\caption{Secrecy rate vs. the y-coordinate of Bob.} 
			\label{result3}
		\end{minipage}
\end{figure*}

\begin{lemma} \label{lemma 1}
	\emph{
	For any UAV horizontal location, $\left\{\hat{\boldsymbol{a}}\right\}$, we have
	\begin{eqnarray}
	\label{16}
	\varphi\left(\boldsymbol{a}\right) & \geq & \varphi^{lb}\left(\boldsymbol{a}\right) \nonumber\\
	& \triangleq & \hat{\tau}\left(\boldsymbol{a}\right)+ \hat{\Lambda}\left(\boldsymbol{a}\right)\left({\lVert\boldsymbol{a}-\boldsymbol{y}\lVert}^2-{\lVert\hat{\boldsymbol{a}}-\boldsymbol{y}\lVert}^2\right), 
	\end{eqnarray}
	where the equality holds at the point $\boldsymbol{a}=\hat{\boldsymbol{a}}$. The coefficients $\hat{\tau}(\boldsymbol{a})$ and $\hat{\Lambda}(\boldsymbol{a})$ are given in Appendix A.
    } 	
\end{lemma}
\begin{IEEEproof} \label{proof 1}
	See Appendix A.
\end{IEEEproof}

Modifying problem (\ref{15}) by replacing $\varphi(\boldsymbol{a},\boldsymbol{n})$ and $\varphi(\boldsymbol{a},\boldsymbol{r})$ with their lower bounds in Lemma {\ref{lemma 1}} yields the following approximated problem
\begin{subequations}\label{17}
	\begin{eqnarray}
	\label{17a}
	&\underset{\boldsymbol{a}}{\rm max}&
	\frac{\left|\varphi^{lb}\left(\boldsymbol{a},\boldsymbol{r}\right)\alpha_{ab}+\varphi^{lb}\left(\boldsymbol{a},\boldsymbol{b}\right)\omega_{ab}\right|^2+1}{\left|\varphi^{lb}\left(\boldsymbol{a},\boldsymbol{r}\right)\alpha_{ae}+\varphi^{lb}\left(\boldsymbol{a},\boldsymbol{e}\right)\omega_{ae}\right|^2+1}  \\
	&\operatorname{s.t.}&R\left(\boldsymbol{a}\right)\geq0,
	\end{eqnarray}
\end{subequations}
where $R(\boldsymbol{a})=|\varphi^{lb}(\boldsymbol{a},\boldsymbol{r})\alpha_{ab}+\varphi^{lb}(\boldsymbol{a},\boldsymbol{b})\omega_{ab}|^2+1-2^{R_\textrm{min}}$.

Similar to problem (\ref{9}), the fractional programming method can be invoked to transform problem (\ref{17}) as follows
\begin{subequations}\label{18}
	\begin{eqnarray}
	\label{18a}
	&\underset{\boldsymbol{a}}{\rm min}&
	\varphi\left(\boldsymbol{a}|\rho\right) \\
	&\operatorname{s.t.}&R\left(\boldsymbol{a}\right)\geq0,
	\end{eqnarray}
\end{subequations}
where $\varphi(\boldsymbol{a}|\rho)=|\varphi^{lb}(\boldsymbol{a},\boldsymbol{r})\alpha_{ae}+\varphi^{lb}(\boldsymbol{a},\boldsymbol{e})\omega_{ae}|^2+1-\rho R(\boldsymbol{a})$, and $\rho\geq0$ is an introduced parameter.

It can be verified that both the subtrahend and minuend of (\ref{18a}) with given $\rho$ are convex w.r.t $\boldsymbol{a}$, thus problem (\ref{18}) can be near-optimally solved by the difference-of-convex (DC) algorithm \cite{DC}. 
After obtaining the optimal solution to problem (\ref{18}), it is not difficult to verify that problem (\ref{17}) can be efficiently solved by the method for solving problem (\ref{9}). 
For space reason, we omit the detailed approximations here, which are similar to those in Section III-B.

Using the results obtained in the previous three subsections, the overall alternating algorithm for solving problem (\ref{5}) is summarized in Algorithm \ref{Algorithm1}, whose convergence and complexity are given in the following proposition.
\newtheorem{Proposition}{\textbf{Proposition}}
\begin{Proposition} \label{proposition 1}
	\emph{
	As long as the number of iterations $L$ is large enough, Algorithm \ref{Algorithm1} is guaranteed to converge, and its computational complexity can be represented by $\mathcal{O}(L\log_2{\frac{1}{\epsilon}})$.
    }
\end{Proposition}
\begin{IEEEproof} \label{proof 2}
	See Appendix B.
\end{IEEEproof}

\section{Simulation Results}
This section presents numerical results to characterize the performance of our proposed algorithm. 
The location of Rose and Eve are respectively set to $\boldsymbol{r}=(3, 5)$ and  $\boldsymbol{e}=(0, 10)$, and the x-coordinate of Bob is set to 0. 
Other simulation parameters are given in Table \ref{table_example}.
Specifically, the BS deployed at $(45, 5, 20)$ is considered as the benchmark for comparison.

Fig.\ref{result1} shows the impact of the transmit power $P_\textrm{max}$ on the secrecy performance with $M=60$ and $\boldsymbol{b}=(0, 0)$. 
A general trend from Fig.\ref{result1} is that the secrecy rates of all the schemes increase with the growth of the transmit power as expected. 
Note that (UAV, IRS) always yields a higher value than the others, while (BS, No-IRS) \cite {PLS-1} looks the worst, which indicates that deploying IRS and UAV is a promising approach for improving wireless communications security. 
And (BS, IRS) \cite{IRS-3} performs better than (UAV, No-IRS) \cite{UAV-1}, which implies that in this challenging setup with both Bob and Eve near to IRS, deploying IRS with 60 reflection elements is more beneficial for improving security than deploying UAV.
\begin{table}
	\renewcommand{\arraystretch}{1.2}
	\caption{Simulation parameters}
	\label{table_example}
	\scalebox{0.97}{
    \centering
	\begin{tabular}{|c|c|}
		\hline  
		\bfseries Parameter&\bfseries Value\\
		\hline
		Height setting&$H=50$ m, $h=5$ m \\
		\hline
		Path loss at 1 m&$\beta_0=-30\textrm{dB}$\\
		\hline
		Path loss factor\cite{parameter1}&$c_{ab}=c_{ae}=3.5$, $c_{ar}=2.2$, $c_{rb}=c_{re}=2.8$\\
		\hline
		Rician factor\cite{parameter2}&$k_\textrm{min}=A_1=0\textrm{dB}$, $k_\textrm{max}=A_1e^{A_2\frac{\pi}{2}}=30\textrm{dB}$\\
		\hline
		Other parameters&$N=4$, $R_\textrm{min}=1\textrm{bit/s/Hz}$, $\sigma_{b}^2=\sigma_{e}^2=-55\textrm{dBm}$\\
		\hline
	\end{tabular}}
\end{table}

Fig.\ref{result2} depicts the secrecy rate versus the number of reflecting elements at IRS (Rose) with $P_\textrm{max}=50$ dBm and $\boldsymbol{b}=(0, 0)$. 
It is observed that the performance gain obtained from IRS is sensitive to the increase in the number of reflecting elements. 
This is due to the fact that the large-scale IRS can provide a strong cascaded channel for legitimate reception but substantially deteriorate the information reception at the eavesdropper.
Interestingly, we notice that when $M=50$, (IRS, BS) and (UAV, No-IRS) perform almost identically, which shows that the secrecy rate enhancements provided by the deployment of software controllable IRS and high mobility UAV are similar under the parameter settings at this point. 

Fig.\ref{result3} demonstrates the effect of the distance between the IRS and the legitimate user on rate performance by varying the y-coordinate of Bob, with $P_\textrm{max}=50$ dBm and $M=60$. 
As can be observed from Fig.\ref{result3}, since the channel gain is a decreasing function of the distance of IRS-Bob link, when Bob is far away from the IRS, deploying IRS is not obviously more efficient than employing UAV for security provisioning. 
However, when Bob approaches the IRS, the reflect beamforming of the IRS can be fully exploited to enhance the desired signal at the legitimate receiver. 
Thus, the secrecy rates achieved by our proposed IRS-aided systems obtain distinct growth correspondingly.

\section{Conclusion}
This letter investigated the secure communication in UAV-enabled networks aided by the IRS.
A non-convex problem was formulated to maximize the secrecy rate by jointly optimizing the transmit power and location of the UAV as well as the phase shifts at the IRS.
To solve this intractable problem efficiently, an AO based algorithm was proposed by invoking the fractional programming and relaxation methods, such as SCA and DC programming.
Finally, simulation results validated the secrecy performance gain of the proposed algorithm over other three baselines.

\section*{APPENDIX A}
Introducing a constant $\alpha<0$, it can be verified that $\varphi(x)=(X+x)^\alpha\sqrt{(k(x)+1)^{-1}}(\sqrt{k(x)}\boldsymbol{g}+\widetilde{\boldsymbol{g}})$, where $k(x)=A_1\exp(A_2\arcsin{(\frac{\bar{h}}{X+x})})$ is convex w.r.t $x\geq-X$. 
Thus, following the SCA technique, for any given $x_0$, we have
\begin{eqnarray}
\label{22}
&\varphi\left(x\right)\geq\varphi\left(x_0\right)+\varphi_x\left(x_0\right)\left({x-x}_0\right), \ \forall x,
\end{eqnarray}
where
\begin{equation}
\varphi(x_0)=\left(X+x_0\right)^\alpha\sqrt{(k(x_0)+1)^{-1}}(\sqrt{k(x_0)}\boldsymbol{g}+\widetilde{\boldsymbol{g}}),
\end{equation}
\begin{equation}
\varphi_x(x_0)=\bar{k}(x_0)\left\{\left[2\alpha+\widetilde{h}(x_0)\bar{k}^2(x_0)\right]\bar{g}(x_0)-\widetilde{h}(x_0)\boldsymbol{g}\right\},
\end{equation}
with 
\begin{equation}
\bar{k}(x)=\frac{1}{2}(X+x)^{\alpha-1}\sqrt{(k(x)+1)^{-1}k(x)}, 
\end{equation}
\begin{equation}
\bar{g}(x)=\boldsymbol{g}+\widetilde{\boldsymbol{g}}\sqrt{k^{-1}(x)},
\widetilde{h}(x)=A_2\bar{h}\sqrt{((X+x)^2-{\bar{h}}^2)^{-1}}.
\end{equation}
Then, considering $X={\lVert\hat{\boldsymbol{a}}-\boldsymbol{y}\lVert}^2+Z^2$, $x_0=0$, and $x={\lVert\boldsymbol{a}-\boldsymbol{y}\lVert}^2-{\lVert\hat{\boldsymbol{a}}-\boldsymbol{y}\lVert}^2$,  Lemma \ref{lemma 1} are derived, where
\begin{equation}
\label{24}
\hat{\tau}(\boldsymbol{a})=({\lVert\hat{\boldsymbol{a}}-\boldsymbol{y}\lVert}^2+Z^2)^\alpha\sqrt{(k(0)+1)^{-1}}(\sqrt{k(0)}\boldsymbol{g}+\widetilde{\boldsymbol{g}}),
\end{equation}
\begin{eqnarray}
\hat{\Lambda}(\boldsymbol{a})= \bar{k}(0)\left\{\left[2\alpha+\widetilde{h}(0)\bar{k}^2(0)\right]\bar{g}(0)-\widetilde{h}(0)\boldsymbol{g}\right\}.
\end{eqnarray}

\section*{APPENDIX B}
To characterize the convergence of the algorithm, we proved that for the $n$-th iteration, the following update rules hold
\begin{subequations}
	\begin{eqnarray}
	\label{25-1}
	R_s^{(n)}&=&R_s(\boldsymbol{f}^{(n)},\boldsymbol{\Theta}^{(n)},\boldsymbol{a}^{(n)}) \\
	&\overset{(a)}{\leq}&R_s(\boldsymbol{f}^{(n+1)},\boldsymbol{\Theta}^{(n)},\boldsymbol{a}^{(n)}) \\
	&\overset{(b)}{\leq}&R_s(\boldsymbol{f}^{(n+1)},\boldsymbol{\Theta}^{(n+1)},\boldsymbol{a}^{(n)}) \\
	&\overset{(c)}{\leq}&R_s(\boldsymbol{f}^{(n+1)},\boldsymbol{\Theta}^{(n+1)},\boldsymbol{a}^{(n+1)})=R_s^{(n+1)}.
	\end{eqnarray}
\end{subequations}
$(a)$ follows that in Step 3 in Algorithm \ref{Algorithm1}, problem (\ref{5}) is solved optimally with the solution $\boldsymbol{f}^{(n+1)}$; 
Let $\varphi_B(\boldsymbol{\theta})$ and $\varphi_E(\boldsymbol{\theta})$ respectively denote the numerator and the denominator of (\ref{9a}), then $(b)$ can be explained by the following inequalities 
\begin{subequations}
\begin{eqnarray}
    \label{25}
	\varphi_E (\widetilde{\boldsymbol{\theta}} ) &-& \mu^\prime\varphi_B (\widetilde{\boldsymbol{\theta}} ) = \varphi (\widetilde{\boldsymbol{\theta}}|\mu^\prime ) \\	
	&\overset{(d)}{=}& f (\widetilde{\boldsymbol{\theta}}| (\mu^\prime,\widetilde{\boldsymbol{\theta}} ) ) \\
	&\overset{(e)}{\geq}& f (\boldsymbol{\theta}^\star(\mu^\prime)| (\mu^\prime,\widetilde{\boldsymbol{\theta}} ) ) \\
	&\overset{(f)}{\geq}& \varphi (\boldsymbol{\theta}^\star (\mu^\prime )|\mu^\prime ) \\
	&\overset{(g)}{=}& \varphi_E (\boldsymbol{\theta}^\star (\mu^\prime ) )-\mu^\prime\varphi_B (\boldsymbol{\theta}^\star (\mu^\prime ) ) = 0.
\end{eqnarray}
\end{subequations}
where $(d)$ is due to $\varphi(\boldsymbol{\theta}|\mu)=f(\boldsymbol{\theta}|(\mu,\boldsymbol{\theta}))$, $(e)$ holds since $f(\boldsymbol{\theta}|\mu,\widetilde{\boldsymbol{\theta}}))$ is minimized when $\widetilde{\boldsymbol{\theta}}=\boldsymbol{\theta}^\star(\mu^\prime)$, $(f)$ is due to the inequality (\ref{11}) and $(g)$ holds from ${\varphi}^\star(\boldsymbol{\theta}^\star(\mu^\prime)|\mu^\prime)=0$. 
Consequently, we obtain  $\frac{\varphi_B(\widetilde{\boldsymbol{\theta}})}{\varphi_E(\widetilde{\boldsymbol{\theta}})}\leq\frac{1}{\mu^\prime}=\frac{\varphi_B(\boldsymbol{\theta}^\star(\mu^\prime))}{\varphi_E(\boldsymbol{\theta}^\star(\mu^\prime))}$, that is, (\ref{9a}) increases over iterations, thus $(b)$ holds;
$(c)$ means that the value of $R_s(\boldsymbol{f}^{(n+1)},\boldsymbol{\Theta}^{(n+1)},\boldsymbol{a}^{(n)})$ can be improved by successively solving the problem (\ref{18}).
Furthermore, the objective value of problem (\ref{5}) is upper bounded by a finite value, Algorithm \ref{Algorithm1} is guaranteed to converge. 

Note that both step 3 and step 4 of Algorithm \ref{Algorithm1} can obtain closed-form solutions, thus the major computation of Algorithm \ref{Algorithm1} lies in step 5 of solving problem (\ref{15}).
Since the considered system has a small user scale and the transmission is completed in a single time slot, the complexity of solving UAV deployment subproblem can be given as $\mathcal{O}(L\log_2{\frac{1}{\epsilon}})$, which completes the proof of Proposition \ref{proposition 1}.


\begin{thebibliography}{99}
     \bibitem{PLS-1}
     Y. {Liu}, Z. {Qin}, and M. {Elkashlan} \textit{et al}, ``Enhancing the physical layer security of non-orthogonal multiple access in large-scale networks,"  \emph{IEEE Trans. Commun.}, vol. 16, no. 3, pp. 1656-1672, Mar. 2017.
     \bibitem{PLS-2}	
     X. {Yue}, Y. {Liu}, and Y. {Yao} \textit{et al}, ``Secure communications in a unified non-orthogonal multiple access framework," \emph{IEEE Trans. Wireless Commun.}, vol. 19, no. 3, pp. 2163-2178, Mar. 2020.
     \bibitem{UAV-1}
     Z. {Li}, M. {Chen} and C. {Pan} \textit{et al}, ``Joint trajectory and communication design for secure UAV networks," \emph{IEEE Commun. Lett.}, vol. 23, no. 4, pp. 636–639, Apr. 2019.
     \bibitem{UAV-2}
     S. {Zhang}, H. {Zhang}, and Q. {He} \textit{et al}, ``Joint trajectory and power optimization for UAV relay networks," \emph{IEEE Commun. Lett.}, vol. 22, no. 1, pp. 161-184, Jan. 2018.
     \bibitem{UAV-3}
     C. {Zhong}, J. {Yao}, and J. {Xu}, ``Secure UAV communication with cooperative jamming and trajectory control," \emph{IEEE Commun. Lett.}, vol. 23, no. 2, pp. 286–289, Feb. 2019.
     \bibitem{UAV-4}
     Y. {Li}, R. {Zhang}, and J. {Zhang} \textit{et al}, ``Cooperative jamming via spectrum sharing for secure UAV communications," \emph{IEEE Wireless Commun. Lett.}, vol. 9, no. 3, pp. 326-330, Mar. 2020. 
     \bibitem{UAV-5}
     Z. {Yang}, W. {Xu} and M. {Shikh-Bahaei}, ``Energy efficient UAV communication with energy harvesting," \emph{IEEE Trans. Veh. Technol.}, vol. 69, no. 2, pp. 1913-1927, Feb. 2020.
     \bibitem{UAV-6}
     Z. {Yang}, C. {Pan} and M. {Shikh-Bahaei} \textit{et al}, ``Joint altitude, beamwidth, location, and bandwidth optimization for UAV-enabled communications," \emph{IEEE Commun. Lett.}, vol. 22, no. 8, pp. 1716-1719, Aug. 2018.
     \bibitem{UAV-7}
     W. {Ni}, H. {Tian}, and S. {Fan} \textit{et al}, ``Optimal transmission control and learning-based trajectory design for UAV-assisted detection and communication," in \emph{Proc. IEEE PIMRC}, London, UK, Aug. 2020, accepted to appear.
     \bibitem{IRS-1}
     H. {Long}, M. {Chen}, and Z. {Yang} \textit{et al}, ``Reflections in the sky: Joint trajectory and passive beamforming design for secure UAV networks with reconfigurable intelligent surface," Jun. 2020. [Online], Available: https://arxiv.org/abs/2005.10559.
     \bibitem{IRS-2}
     X. {Yu}, D. {Xu}, and R. {Schober}, ``Enabling secure wireless communications via intelligent reflecting surfaces," in \emph{Proc. IEEE GLOBECOM Workshops}, Waikoloa, HI, USA, Dec. 2019.
     \bibitem{challenge}
     Y. {Liu}, X. {Liu}, and X. {Mu} \textit{et al}, ``Reconfigurable intelligent surfaces: Principles and opportunities," Jul. 2020. [Online], Available: https://arxiv.org/abs/2007.03435.
     \bibitem{8743496}
     H. {Shen}, W. {Xu}, and S. {Gong} \textit{et al}, ``Secrecy rate maximization for intelligent reflecting surface assisted multi-antenna communications," \emph{IEEE Commun. Lett.}, vol. 23, no. 9, pp. 1488-1492, Sept. 2019.
     \bibitem{CSI_1} 
     J. {Chen}, Y. C. {Liang}, and H. V. {Cheng} \textit{et al}, ``Channel estimation for reconfigurable intelligent surface aided multiuser MIMO systems," Dec. 2019. [Online], Available: https://arxiv.org/abs/1912.03619.
     \bibitem{CSI_2}
     L. {Wei}, C. {Huang}, and G. C. {Alexandropoulos} \textit{et al}, ``Channel estimation for RIS-empowered multi-user MISO wireless communications," Aug. 2020. [Online], Available: https://arxiv.org/abs/2008.01459.
     \bibitem{IRS-3}
     L. {Dong} and H. {Wang}, ``Secure MIMO transmission via intelligent reflecting surface," \emph{IEEE Wireless Commun. Lett.}, vol. 9, no. 6, pp. 787-790, Jun. 2020. 
     \bibitem{IRS}
     A. {Almohamad}, A. {Tahir}, and A. {Al-Kababji} \textit{et al}, ``Smart and secure wireless communications via reflecting intelligent surfaces: A short survey," Sept. 2020. [Online], Available: https://arxiv.org/abs/2006.14519v4.
     \bibitem{Werner1967On}
     W. {Dinkelbach}, ``On nonlinear fractional programming," \emph{Management Science}, vol. 13, no. 7, pp. 492-498, 1967.
     \bibitem{7547360}
     Y. {Sun}, P. {Babu}, and D. P. {Palomar}, ``Majorization-minimization algorithms in signal processing, communications, and machine learning," \emph{IEEE Trans. Signal Process.}, vol. 65, no. 3, pp. 794-816, Feb. 2017.
     \bibitem{DC}
     R. {Horst} and N. V. {Thoai}, ``DC programming: Overview", \emph{J. Optim. Theory Appl.}, vol. 103, no. 1, pp. 1-43, Oct. 1999.
    \bibitem{parameter1}
    T. {Jiang} and Y. {Shi}, ``Over-the-air computation via intelligent reflecting surfaces," in \emph{Proc. IEEE GLOBECOM Workshops}, Waikoloa, HI, USA, Dec. 2019.
    \bibitem{parameter2}
    Iskandar and S. {Shimamoto}, ``The channel characterization and performance evaluation of mobile communication employing stratospheric platform," in \emph{Proc. IEEE/ACES Int. Conf. Wireless Commun. Appl. Comput. Electromag.}, Honolulu, HI, USA, Apr. 2005.
\end{thebibliography}
\end{document}